\documentclass[graybox]{svmult}

\usepackage{mathptmx}       
\usepackage{helvet}         
\usepackage{courier}        
\usepackage{type1cm}        
\usepackage{makeidx}         
\usepackage{graphicx}        
\usepackage{multicol}        
\usepackage[bottom]{footmisc}

\usepackage{amssymb}
\usepackage{amsmath}
\usepackage{amsfonts}
\usepackage{color}
\usepackage{epstopdf}

\DeclareMathOperator{\sign}{\operatorname{sign}}

\makeindex             

\begin{document}

\title*{Measures of spike train synchrony and directionality}

\author{Eero Satuvuori and Irene Malvestio and Thomas Kreuz}
\institute{Eero Satuvuori \at Department of Physics and Astronomy,
University of Florence, Sesto Fiorentino, Italy\\
Faculty of Behavioural and Movement Sciences ,Vrije Universiteit Amsterdam, Netherlands\\ \email{eero.satuvuori@unifi.it}
\and Irene Malvestio \at Department of Information and Communication Technologies, 
Universitat Pompeu Fabra, Barcelona, Spain \\
Department of Physics and Astronomy, University of Florence, Sesto Fiorentino, Italy\\ \email{irene.malvestio@upf.edu}
\and Thomas Kreuz \at Institute for Complex Systems, CNR, Sesto Fiorentino, Italy\\ \email{thomas.kreuz@cnr.it }}
%
%
\maketitle

\abstract*{
Measures of spike train synchrony have become important tools in both experimental
and theoretical neuroscience.
Three time-resolved measures called the ISI-distance, the SPIKE-distance, and
SPIKE-synchronization have already been successfully applied in many different
contexts.
These measures are time scale independent, since they consider all time scales
as equally important.
However, in real data one is typically less interested in the smallest time scales
and a more adaptive approach is needed.
Therefore, in the first part of this Chapter we describe recently introduced
generalizations of the three measures, that gradually disregard differences in
smaller time-scales. 
Besides similarity, another very relevant property of spike trains is the temporal
order of spikes. 
In the second part of this chapter we address this property and describe a very recently
proposed algorithm, which quantifies the directionality within a set of spike train.
This multivariate approach sorts multiple spike trains from leader to follower and
quantifies the consistency of the propagation patterns.
Finally, all measures described in this chapter are freely available for download.
}

\abstract{
Measures of spike train synchrony have become important tools in both experimental
and theoretical neuroscience.
Three time-resolved measures called the ISI-distance, the SPIKE-distance, and
SPIKE-synchronization have already been successfully applied in many different
contexts.
These measures are time scale independent, since they consider all time scales
as equally important.
However, in real data one is typically less interested in the smallest time scales
and a more adaptive approach is needed.
Therefore, in the first part of this Chapter we describe recently introduced
generalizations of the three measures, that gradually disregard differences in
smaller time-scales. 
Besides similarity, another very relevant property of spike trains is the temporal
order of spikes. 
In the second part of this chapter we address this property and describe a very recently
proposed algorithm, which quantifies the directionality within a set of spike train.
This multivariate approach sorts multiple spike trains from leader to follower and
quantifies the consistency of the propagation patterns.
Finally, all measures described in this chapter are freely available for download.
}

\section{Introduction}
\label{sec:1}

The brain can be considered as a huge network of spiking neurons.
It is typically assumed that only the spikes, and not the shape of the action
potential nor the background activity, convey the information processed within
this network \cite{Kreuz13c}.
Sequences of consecutive spikes are called spike trains.
Measures of spike train synchrony are estimators of the similarity between two
or more spike trains, which are important tools for many applications in neuroscience. 
Among others, they allow to test the performance of neuronal models \cite{Jolivet08},
they can be used to quantify the reliability of neuronal responses upon repeated
presentations of a stimulus \cite{Mainen95}, and they help in the understanding
of neural networks and neural coding \cite{Victor05}.

Over the years many different methods have been developed in order to quantify
spike train synchrony.
They can be divided in two classes: time-scale dependent and time-scale independent methods. 
The two most known time-scale dependent methods are the Victor-Purpura  distance \cite{Victor96}
and the  van Rossum  distance  \cite{VanRossum01}.
They  describe spike train (dis)similarity based on a user-given time-scale to which
the measures are mainly sensitive to.
Time scale independent methods have been developed more recently. 
In particular, the ISI-distance \cite{Kreuz07a}, the SPIKE-distance \cite{Kreuz11, Kreuz13}
and SPIKE-synchronization \cite{Kreuz15} are parameter-free distances, with the capability
of discerning similarity across different spatial scales.
All of these measures are time-resolved, so they are able to analyze the time dependence
of spike train similarity.

One problematic aspect of time-scale independent methods is that they consider all
time-scales as equally important.
However, in real data one typically is not interested in the very small time scales.
Especially in the presence of bursts (multiple spikes emitted in rapid succession),
a more adaptive approach that gradually disregards differences in smaller time-scales
is needed.
Thus, in the first part of this chapter we describe the recently developed adaptive
extensions of these three parameter-free distances: A-ISI-distance, A-SPIKE-distance
and A-SPIKE-synchronization \cite{Satuvuori17}. 

All of these  similarity measures are symmetric and in consequence invariant to changes
in the order of spike trains. 
However, often information about directionality is needed, in particular in the study
of propagation phenomena.
For example, in epilepsy studies, the analysis of the varying similarity patterns of
simultaneously recorded ensembles of neurons can lead to a better understanding of the
mechanisms of seizure generation, propagation, and termination \cite{Truccolo11, Bower12}.

In the second part of this chapter we address the question: Which are the neurons
that tend to fire first, and which are the ones that tend to fire last?
We present SPIKE-Order \cite{Kreuz2016}, a recently developed algorithm which is able
to discern propagation pattern in neuronal data.
It is a multivariate approach which allows to sort multiple spike trains from leader
to follower and to quantify the consistency of the temporal leader-follower relationships.
We close this chapter by describing some applications of the methods presented.

\section{Measures of spike train synchrony}
\label{sec:2}

Two of the most well known spike train distances, the Victor-Purpura \cite{Victor96}
and the van Rossum  distance \cite{VanRossum01}, are time-scale dependent.
One drawback of these methods is the fixed time-scale, since it sets
a boundary between rate and time coding for the whole recording.
In the presence of bursts, where multiple spikes are emitted in rapid
succession, there are usually many time-scales in the data and this is difficult
to detect when using a measure that is sensitive to only one time-scale at a time
\cite{Chicharro11}. 

The problem of having to choose one time-scale has been eliminated in the time-scale
independent ISI-distance \cite{Kreuz07a}, SPIKE-distance \cite{Kreuz11, Kreuz13}
and SPIKE-synchronization \cite{Kreuz15},
since these methods always adapt to the local firing rate.
The ISI-distance and the SPIKE-distance are time resolved, time-scale free measures of
dissimilarity between two or more spike trains.
The ISI-distance is a measure of rate dissimilarity.
It uses the interspike intervals (ISIs) to estimate local firing rate of spike trains
and measures time-resolved differences between them.
The SPIKE-distance, on the other hand, compares spike time
accuracy between the spike trains and uses local firing rates to adapt to the
time-scale.
SPIKE-synchronization is also time-scale free and is a discrete
time resolved measure of similarity based on ISI derived coincidence windows
that determine if two spikes in a spike train set are coincident or not. 

The ISI-distance, SPIKE-distance, and SPIKE-synchronization are looking at all time-scales at the same time. 
However, in real data not all time-scales are equally important, 
and this can lead to spuriously high values of dissimilarity when looking only at the local information. 
Many sequences of discrete events contain different time-scales.
For example, in neuronal recordings besides regular spiking one often finds bursts,
i.e., rapid successions of many spikes.
The A-ISI-distance, A-SPIKE-distance and A-SPIKE-synchronization \cite{Satuvuori17}
are generalized versions of previously published methods the ISI-distance
\cite{Kreuz07c}, SPIKE-distance \cite{Kreuz11} and SPIKE-synchronization \cite{Kreuz15}. 
The generalized measures also contain a notion of global context that discriminates
between relative importance of differences in the global scale.
This is done by means of a normalization based on a minimum relevant time-scale (MRTS).
They start to gradually ignore differences between spike trains for interspike
intervals (ISIs) that are smaller than the MRTS.
The generalization provided by the MRTS is implemented with the threshold parameter $thr$, which is then applied in a different way to each of the measures.
The threshold is used to determine if a difference between the spike trains should
be assessed in a local or in a global context.
This threshold is used for all three measures, but the way it is applied varies.
The extended methods fall back to the original definitions when $thr = 0$ and we
refer to this whenever we talk of the original methods.
In this case even the smallest time-scales matter and all differences are assessed
in relation to their local context only.

Throughout this Section we denote the number of spike trains by $N$, indices of
spike trains by $n$ and $m$, spike indexes by $i$ and $j$ and the number of spikes
in spike train $n$ by $M_n$.
The spike times of spike train $n$ are denoted by $\{t^{(n)}_i\}$ with $i=1\dots
M_n$.

\begin{figure}
	\includegraphics[width = 0.8\textwidth]{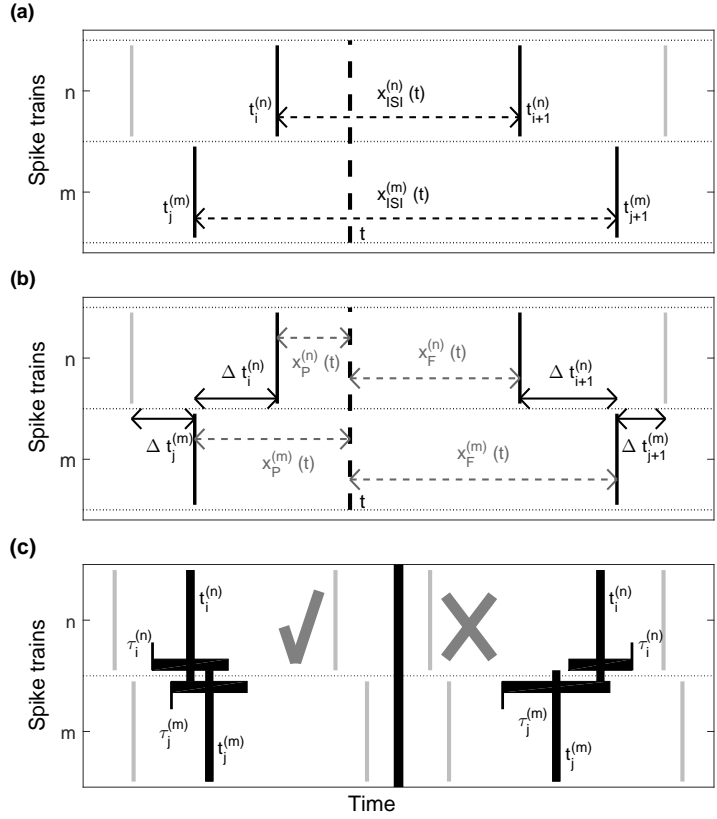}
	\centering
	\caption[Figure]{
		Schematic drawing for all three measures. 
		(a) Illustration of the variables used to define the \textbf{A-ISI-distance}.
		All measures use the instantaneous interspike interval $x_{ISI}^{(n)}(t)$ to
		adapt to the local firing rate.
		(b) Additional variables used for the \textbf{A-SPIKE-distance}.
		(c) Coincidence criterion for the \textbf{A-SPIKE-synchronization}.
		The coincidence window of each spike is derived from the two surrounding
		interspike intervals.
		For simplicity the $thr = 0$ case is shown.
		Here we illustrate two different examples.
		The two spikes on the left side are considered coincident since both lie in each
		other's coincidence windows.
		On the right there is no coincidence because the spike from the second spike train
		is outside of the coincidence window from the spike of the first spike train.
		\label{Fig:Illustration}}
\end{figure}

\subsection{\label{ss:ISI-dist}Adaptive ISI-distance}

The A-ISI-distance \cite{Satuvuori17} measures the instantaneous rate difference
between spike trains (see Fig. \ref{Fig:Illustration}a).
It relies on a time-resolved profile, meaning that in a first step a dissimilarity
value is assigned to each time instant.
To obtain this profile, we first assign to each time instant $t$ the times of the
previous spike and the following spike

\begin{eqnarray}
t_{\mathrm {P}}^{(n)} (t) = \max\{t_i^{(n)} | t_i^{(n)} \leq t\}  \quad
\textrm{for }\quad t_1^{(n)} \leqslant t \leqslant t_{M_n}^{(n)} \label{eq:Prev-Spike}\\
t_{\mathrm {F}}^{(n)} (t) = \min\{t_i^{(n)} | t_i^{(n)} > t\} \quad \textrm{for } \quad
t_1^{(n)} \leqslant t \leqslant t_{M_n}^{(n)}.\label{eq:Foll-Spike}
\end{eqnarray}
From this for each spike train $n$ an instantaneous ISI can be calculated as
\begin{equation} \label{eq:ISI}
x_{\mathrm{ISI}}^{(n)} (t) = t_{\mathrm {F}}^{(n)} (t) - t_{\mathrm {P}}^{(n)} (t).
\end{equation}
The A-ISI-profile is defined as a normalized instantaneous ratio in ISIs:
\begin{equation} \label{eq:pairwise A-ISI-profile}
I_{n,m}^A (t) = \frac{|x_{\mathrm{ISI}}^{(n)} (t) - x_{\mathrm{ISI}}^{(m)}
	(t)|}{\max \{x_{\mathrm{ISI}}^{(n)} (t), x_{\mathrm{ISI}}^{(m)} (t),thr\} }.
\end{equation}
For the A-ISI-distance the MRTS is defined so that when the ISI of both spike trains
are smaller than a threshold value $thr$, the threshold value is used instead.
The multivariate A-ISI-profile is obtained by averaging over all pairwise A-ISI-profiles:

\begin{figure}[t]
	\centering
	\includegraphics[width = 0.8\textwidth]{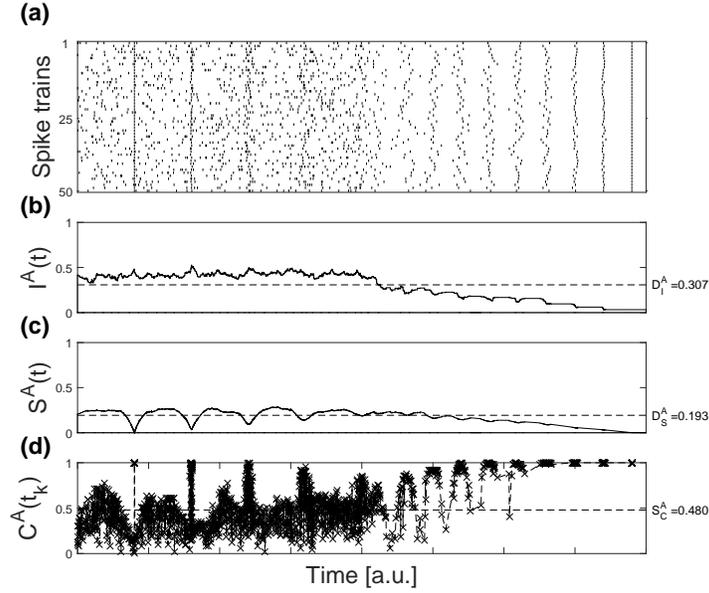}
	\caption[Figure]{
		Profiles of \textbf{A-ISI-distance} (a), \textbf{A-SPIKE-distance}
		(b) and \textbf{A-SPIKE-synchronization} (c) for an artificial example dataset
		of 50 spike trains with population events with different jitters and decreasing noise over time.
		\label{example_A_spikes}}	
\end{figure}
\begin{equation} \label{eq:multivariate ISI}
I^A (t) = \frac{1}{N(N-1)/2}\sum_{n=1}^{N-1} \sum_{m=n+1}^N I_{n,m}^A (t).
\end{equation}
This is a non-continuous piecewise constant profile and integrating over time
gives the A-ISI-distance: 
\begin{equation} \label{eq:ISI-distance}
D_I^A = \frac{1}{t_e-t_s} \int_{t_s}^{t_e} I^A (t)dt .
\end{equation}
Where $t_s$ and $t_e$ are the start and end times of the recording respectively. If thr is set to zero, the method falls back to the ISI-distance \cite{Kreuz07c}. 

Fig. \ref{example_A_spikes}a shows an artificial spike train dataset together
with the corresponding A-ISI-profile in Fig. \ref{example_A_spikes}b.
The A-ISI-profile for the example dataset shows high dissimilarity for the left
side of the raster plot, where noise is high.
When the noise is decreased and rates become more similar in the right side, the
dissimilarity profile goes down.
The overall ISI-distance is the mean value of the profile.

\subsection{\label{ss:SPIKE-dist}Adaptive SPIKE-distance}

The A-SPIKE-distance \cite{Satuvuori17} measures the accuracy of spike
times between spike trains relative to local firing rates (see Fig. \ref{Fig:Illustration}b).
In order to assess the accuracy of spike events, each spike is assigned a
distance to its nearest neighbour in the other spike train:
\begin{equation} \label{eq:Closest spike}
\Delta t_i^{(n)} = \min_j(|t_i^{(n)} - t_j^{(m)}|).
\end{equation}
The distances are interpolated between spikes using for all times $t$ the time differences to
the previous and to the following spikes ${x_P^{(n)}(t)}$ and ${x_F^{(n)}(t)}$:

\begin{eqnarray}
x_P^{(n)}(t) = t-t_i^{(n)}    \quad \textrm{for }\quad  t_i^{(n)} \leqslant t \leqslant
t_{i+1}^{(n)}\\
x_F^{(n)}(t) = t_{i+1}^{(n)}-t \quad \textrm{for }\quad  t_i^{(n)} \leqslant t \leqslant
t_{i+1}^{(n)}.
\end{eqnarray}
These equations provide time-resolved quantities needed to define time-resolved dissimilarity
profile from discrete values the same way as Eqs. \ref{eq:Prev-Spike} and \ref{eq:Foll-Spike}
provide them for A-ISI-distance.
The weighted spike time difference for a spike train is then calculated as an
interpolation from one difference to the next by
\begin{equation} \label{eq:weighted distance}
S_{n} (t) = \frac{\Delta t_i^{(n)} (t)x_F^{(n)}(t) + \Delta t_{i+1}^{(n)}
	(t)x_P^{(n)}(t)}{x_{\mathrm{ISI}}^{(n)} (t)}\quad,\quad  t_i^{(n)} \leqslant t \leqslant
t_{i+1}^{(n)}.
\end{equation}
This continuous function is analogous to term ${x_{\mathrm{ISI}}^{(n)}}$ for the ISI-distance,
except that it is piecewise linear instead of piecewise constant.
The pairwise A-SPIKE-distance profile is calculated by temporally averaging the weighted
spike time differences, normalizing to the local firing rate average and, finally,
weighting each profile by the instantaneous firing rates of the two spike trains:

\begin{figure}
	\centering
	\includegraphics[width = 0.8\textwidth]{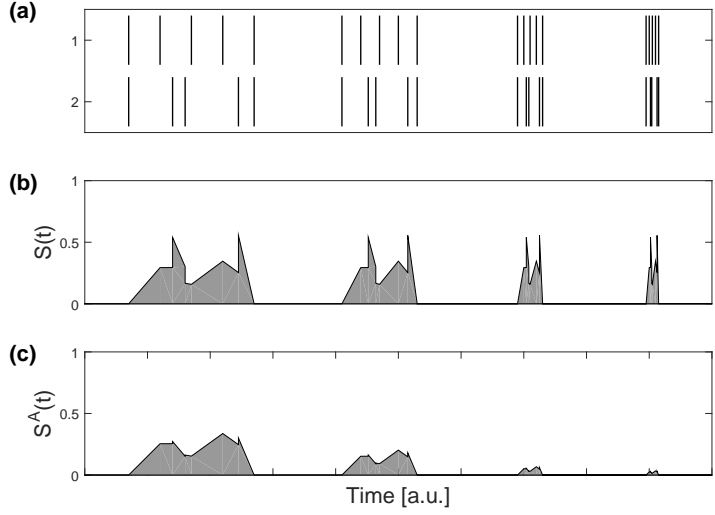}
	\caption[Figure]{
		An example spike train pair and its SPIKE-distance and A-SPIKE-distance
		profiles.
		(a) Two spike trains having four events with five spikes per event in each spike train.
		The sequence of spikes in all four events is the same but the event is increasingly compressed.
		The only thing that changes is the time-scale.
		From a global perspective the first event consists of non-synchronous
		individual spikes, while the last event consists of coincident bursts.
		The two events in the middle are intermediates.
		(b) The SPIKE-distance $S(t)$ looks only at the local context and has the same profile
		shape for all events.
		(c) The A-SPIKE-distance considers also the global context and judges the first event
		like the SPIKE-distance as being dissimilar, but scales down the small spike time
		differences in the burst and considers the coincident burst as very similar.
		\label{motivation3}}	
\end{figure}

\begin{equation} \label{eq:A-SPIKE profile}
S_{m,n}^{A} (t) = \frac{S_n x_{\mathrm{ISI}}^m (t) +S_m x_{\mathrm{ISI}}^n (t) }{2\langle
	x_{\mathrm{ISI}}^{n,m} (t)\rangle \max\{ \langle x_{\mathrm{ISI}}^{n,m} (t)\rangle, thr \} },
\end{equation}
where $\langle x_{\mathrm{ISI}}^{n,m} (t)\rangle$ is the mean over the two instantaneous ISIs. MRTS is defined by using a threshold, that replaces the denominator of weighting to
spike time differences if the mean is smaller than the $thr$.
This profile is analogous to the pairwise A-ISI-profile ${I_{n,m}^A (t)}$, but again it is 
piecewise linear, not piecewise constant.
Unlike $S_{n} (t)$ it is not continuous, typically it exhibits instantaneous jumps at the times
of the spikes.
The multivariate A-SPIKE-profile is obtained the same way as the multivariate
A-ISI-profile, by averaging over all pairwise profiles:
\begin{equation} \label{eq:SPIKE profile}
S^A (t) = \frac{1}{N(N-1)/2}\sum_{n=1}^{N-1} \sum_{m=n+1}^N S_{m,n}^{A} (t).
\end{equation}
The final A-SPIKE-distance is calculated as the time integral over the multivariate
A-SPIKE-profile the same way as the A-ISI-distance:
\begin{equation} \label{eq:Spike-Distances}
D_S^A = \frac{1}{t_e-t_s} \int_{t_s}^{t_e} S^A(t)dt.
\end{equation}
The effect of applying the threshold can be seen in Fig. \ref{motivation3}.
With $thr = 0$ the method falls back to the regular SPIKE-distance \cite{Kreuz11}.
The A-SPIKE-profile for the artificial test dataset in Fig. \ref{example_A_spikes}c
goes to zero when the spikes in all spike trains appear at the exactly same time.

\subsection{\label{ss:SPIKE-sync}Adaptive SPIKE-synchronization}

A-SPIKE-synchronization \cite{Satuvuori17} quantifies how many of the
possible coincidences in a dataset are actually coincidences (Fig. \ref{Fig:Illustration}c).
While the A-ISI-distance and the A-SPIKE-distance are measures of dissimilarity which
obtain low values for similar spike trains, A-SPIKE-synchronization measures similarity.
If all the spikes are coincident with a spike in all the other spike trains, the
value will be one.
In contrast, if none of the spikes are coincident, it will be zero.

The original SPIKE-synchronization \cite{Kreuz15} is parameter- and time-scale-free,
since it uses the adaptive coincidence detection which was first proposed for the
measure Event synchronization \cite{QuianQuiroga02b}. 
The coincidence window, i.e., the time lag below which two spikes from two different
spike trains, $t_i^{(n)}$ and $t_j^{(m)}$, are considered to be coincident, is adapted
to the local firing rate.
Spikes are coincident only if they both lie in each others coincidence windows.
A-SPIKE-synchronization is a generalized version of the SPIKE-synchronization.
The MRTS is used to decide if the window is determined locally or if the global context
should be taken into account.

As a first step, we define the ISI before and after the spike as

\begin{eqnarray}
x_{iP}^{(n)}= \lim_{t \to t_i-} x_{\mathrm{ISI}}^{(n)} (t)\\
x_{iF}^{(n)}= \lim_{t \to t_i+} x_{\mathrm{ISI}}^{(n)} (t).
\end{eqnarray}
The coincidence window for spike $i$ of spike train $n$ is defined by determining a
minimum coincidence window size for a spike as half of the ISIs adjacent to the spike and allowing asymmetric coincidence windows based on MRTS.
This is done by using $thr$ instead of the minimum, if it is smaller.
Since the threshold value is based on ISIs and the coincidence window spans both
sides of the spike, only half of the threshold spans each side.
For the A-ISI- and the A-SPIKE-distance the changes induced by the threshold appear
gradually, but for A-SPIKE-synchronization it is a sudden change from a non-coincidence to coincidence for a spike.
Therefore, due to the binary nature of A-SPIKE-synchronization, the threshold is additionally divided by two.
The coincidence window is not allowed to overlap with a coincidence window of
another spike and is thus limited to half the ISI even if the threshold is
larger.
The base of the window is defined by the two adjacent ISIs:

\begin{equation}\label{eq:A-councidence}
\tau_{i}^{(n)} = \frac{1}{2}\min \{ x_{iP}^{(n)},	x_{iF}^{(n)} \}.
\end{equation}
The coincidence window of a spike is then defined in an asymmetric form by using the
coincidence window part before and after the spike as

\begin{eqnarray}
\tau_{iP}^{(n)} = \min \{ \max (\frac{1}{4}thr,
\tau_{i}^{(n)}),\frac{1}{2}x_{iP}^{(n)} \} \label{eq:A-CoincidenceP}\\
\tau_{iF}^{(n)} = \min \{ \max (\frac{1}{4}thr,
\tau_{i}^{(n)}),\frac{1}{2}x_{iF}^{(n)} \}. \label{eq:A-CoincidenceF}
\end{eqnarray}
The combined coincidence window for spikes $i$ and $j$ is then defined as

\begin{equation} \label{eq:A-Coincidence-MaxDist}
\tau_{ij}^{(n,m)} = \begin{cases}
\min \{\tau_{iF}^{(n)},\tau_{jP}^{(m)}\}    & {\rm if} ~~ t_i \leqslant t_j \cr
\min \{\tau_{iP}^{(n)},\tau_{jF}^{(m)}\}      &  otherwise
\end{cases}.
\end{equation}
The coincidence criterion can be quantified by means of a coincidence indicator
\begin{equation} \label{eq:SPIKE-Coincidence}
C_i^{(n,m)} = \begin{cases}
1     & {\rm if} ~~ \min_j \{|t_i^{(n)} - t_j^{(m)}|\} < \tau_{ij}^{(n,m)} \cr
0     & {\rm otherwise}
\end{cases}.
\end{equation}
This definition ensures that each spike can only be coincident with at most one
spike in the other spike train.
The coincidence criterion assigns either a one or a zero to each spike depending
on whether it is part of a coincidence or not.
For each spike of every spike train, a normalized coincidence counter
\begin{equation} \label{eq:Multi-SPIKE-Coincidence}
C_i^{(n)} = \frac1{N-1}\sum_{m\neq n} C_i^{(n,m)}
\end{equation}
is obtained by averaging over all $N-1$ bivariate coincidence indicators
involving the spike $i$ in spike train $n$.

This way we have defined a coincidence indicator for each individual spike in
the spike trains.
In order to obtain one combined similarity profile, we pool the spikes of the
spike trains as well as their coincidence indicators by introducing one overall
spike index $k$ and defining
\begin{figure}
	\centering
	\includegraphics[width = 0.8\textwidth]{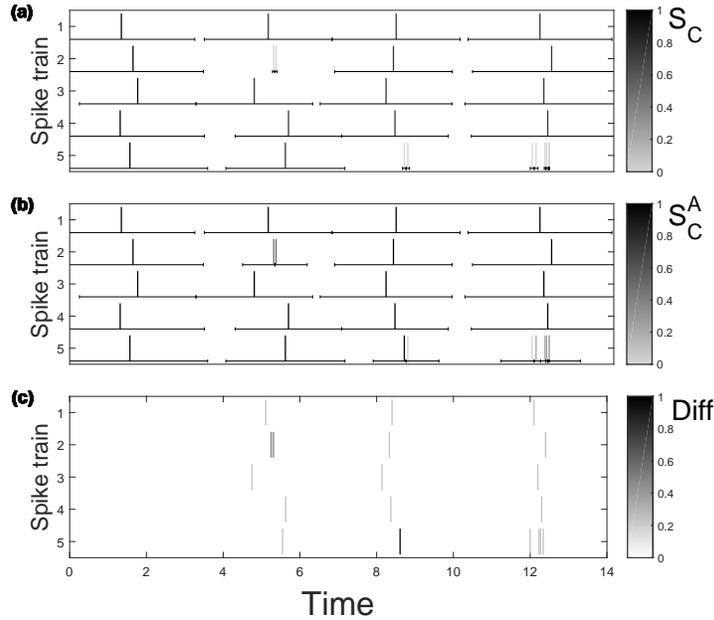}
	\caption{
		SPIKE-synchronization, A-SPIKE-synchronization and their difference
		illustrated using five spike trains with four simple events. 		
		Without the correction (A) in case of spike doublets (first and second event)
		or bursts (third event) the small interspike intervals result in an unreasonably
		high demand for spike timing accuracy.
		With the adaptive correction (B) for all these cases the likelihood increases
		that at least one of the spikes is part of a coincidence.
		On the other hand, if there are no doublets or bursts (last event), nothing changes
		(best seen in C).
		Note that for better visibility the colour scales differ, we use grey-black in A and B and
		white-black in C. 
		\label{fig:A-SYNC_simple}}
\end{figure}
\begin{equation} \label{eq:Sum_of_all_spikes}
M = \sum_{n=1}^N M_n.
\end{equation}
This yields one unified set of coincidence indicators $C_k$:
\begin{equation} \label{eq:Multi-Profile}
\{C_k\} = \bigcup_n \{C_i^{(n)} \}.
\end{equation}
From this discrete set of coincidence indicators $C_k$ the A-SPIKE-synchronization
profile $C^A (t_k)$ is obtained by $C^A (t_k) = C_k$.

Finally, A-SPIKE-synchronization is defined as the average value of this profile
\begin{equation} \label{eq:Multi-SPIKE-Synchronization}
S_C^{A} = \frac{1}{M} \sum_{k=1}^M C^A (t_k).
\end{equation}
It is important to note that since A-SPIKE-synchronization is a measure of similarity,
reducing differences below threshold adds coincidences and thus the value obtained increases.
In Fig. \ref{fig:A-SYNC_simple} we illustrate how the asymmetric coincidence
windows of A-SPIKE-synchronization allow for a better coverage of burst events
which makes it easier to match spikes when compared to SPIKE-synchronization ($thr = 0$) \cite{Kreuz15}. 

As can be seen in Fig. \ref{example_A_spikes}d, the A-SPIKE-synchronization profile
is discrete and only defined at spike times.
A dotted line between the points is added as visual aid.
The profile gets higher values the more coincidences are found for each spike in other spike trains.

\subsection{\label{ss:MRTS_thr} Selecting the threshold value}

In some cases spikes that occur less than a second apart might be considered
more simultaneous than those taking place within minutes, and in applications like
meteorological systems, weeks instead of months.
Setting the minimum relevant time-scale might not be a simple task. 
If no information of the system producing the spikes is available, all one can do to estimate
an appropriate threshold value is to look at the ISIs.

There are two criteria that a threshold value extracted from the data has to fulfil. 
First of all it needs to adapt to changes in spike count so that adding more spikes
gives shorter threshold.
Additionally we want the threshold to adapt to changes in the ISI-distribution when
the spike count is fixed.
The more pronounced bursts are found in the data, the more likely any differences
within aligned bursts are not as important as their placement.
Thus, we want our threshold to get longer if the spikes are packed together.
To do so, all the ISIs in all spike trains are pooled and the threshold is determined
from the pooled ISI distribution.

One should not just take a value based on ISI-distribution that counts the interspike
intervals, as the mean does, but weight them by their length, which is equivalent to
taking the average of the second moments of ISIs. 
Doing this reduces the importance of very short ISIs even if they are statistically much more common. 
In order to obtain a value with the right dimension, the square root of
the average must be taken:

\begin{equation} \label{eq:distribution average ^2}
thr = \sqrt{\langle {(L_{\mathrm{ISI}})^2}\rangle} = \sqrt{\frac{\sum_{n=1}^{N} a_n
		{L_{\mathrm{ISI}}^n}^2 }{\sum_{n=1}^{N}a_n}}.
\end{equation}
Here we denoted a single ISI length in the pooled distribution as $L_{\mathrm{ISI}}^n$
and the number of ISI with length $L_{\mathrm{ISI}}^n$ as $a_n $.
It is important to note, however, that this is only an estimate based on
different time-scales found in the data.
The selected MRTS is not an indicator of a time-scale of the system that
produced the spikes.

As an example of how the threshold works we apply the threshold to Gamma $\Gamma(k,x)$ distribution. Since the kurtosis of the distribution is proportional to $1/k$, for small k the distribution contains large number of small ISI and few long ones.This is the property the threshold is tracking. The mean of a gamma distribution is $k/x$ and the second moment $(k+1)k/x$. thus the ratio of the threshold and the mean ISI is $thr/\langle {L_{\mathrm{ISI}}}\rangle = \sqrt{x(k+1)/k}$. From the formula we can see that for small k, where the distribution is more skewed, the ratio between the mean ISI and the threshold increases. This means that mainly the rare and large inter-burst ISIs are taken into account.

The threshold value determines the outcome of the adaptive methods.
However, the threshold is not a hard set limiter neglecting everything below the
threshold, but rather the point from which on differences are considered
in a global instead of a local context.


\section{Measures of spike train directionality}
\label{sec:3}

Often a set of spike trains exhibits well-defined patterns of spatio-temporal propagation
where some prominent feature first appears at a specific location and then spreads
to other areas until potentially becoming a global event.
If a set of spike trains exhibits perfectly consistent repetitions of the same global
propagation pattern, this can be called a \textit{synfire pattern}.
For any spike train set exhibiting propagation patterns the questions arises 
naturally whether these patterns show any consistency, i.e., to what extent do 
the spike trains resemble a synfire pattern, are there spike trains that 
consistently lead global events and are there other spike trains that 
invariably follow these leaders?
%
%
\begin{figure}
    \includegraphics[width=85mm]{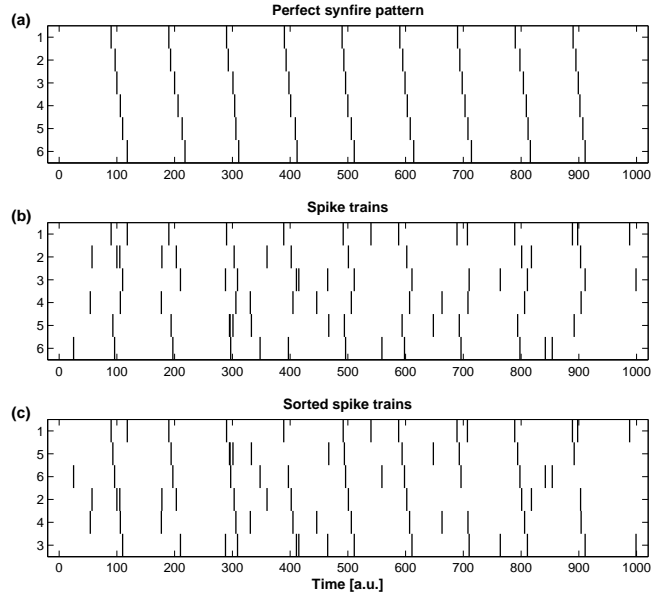}
	\centering
	\caption{\label{fig:SPIKE_Order_Motivation}
	Motivation for SPIKE-order and Spike Train Order.
	(a) Perfect Synfire pattern.
	(b) Unsorted set of spike trains.
	(c) The same spike trains as in (b) but now sorted from leader to follower.}
\end{figure}
%
%

In the second part of this chapter we describe a framework consisting of two
directional measures (\textit{SPIKE-Order} and \textit{Spike Train Order}) that
allows to define a value termed \textit{Synfire Indicator} which quantifies the
consistency of the leader-follower relationships \cite{Kreuz2016}.
This Synfire Indicator attains its maximal value of $1$ for a perfect synfire
pattern in which all neurons fire repeatedly in a consistent order from leader
to follower (Fig. \ref{fig:SPIKE_Order_Motivation}a).

The same framework also allows to sort multiple spike trains from leader to
follower, as illustrated in Figs. \ref{fig:SPIKE_Order_Motivation}b and
\ref{fig:SPIKE_Order_Motivation}c.
This is meant purely in the sense of temporal sequence. Whereas Fig.
\ref{fig:SPIKE_Order_Motivation}b shows an artificially created but rather
realistic spike train set, in Fig. \ref{fig:SPIKE_Order_Motivation}c the
same spike trains have been sorted to become as close as possible to a synfire
pattern.
Now the spike trains that tend to fire first are on top whereas spike trains
with predominantly trailing spikes are at the bottom. 

Analyzing leader-follower relationships in a spike train set requires a
criterion that determines which spikes should be compared against each other.
What is needed is a match maker, a method which pairs spikes in such a way
that each spike is matched with at most one spike in each of the other spike
trains.
This match maker already exists.
It is the adaptive coincidence detection first used as the fundamental
ingredient for the bivariate measure \textit{SPIKE-synchronization}
\cite{Kreuz15} (see Section \ref{ss:SPIKE-sync}).

\subsection{\label{ss:SPIKE-Order} SPIKE-Order and Spike Train Order}

The symmetric measure SPIKE-Synchronization (introduced in Section \ref{ss:SPIKE-sync})
assigns to each spike of a given spike train pair a bivariate coincidence indicator.
These coincidence indicators $C_i^{(n,m)}$ (Eq. \ref{eq:SPIKE-Coincidence}),
which are either $0$ or $1$, are then averaged over spike train pairs and converted
into one overall profile $C(t_k)$ normalized between $0$ and $1$.
In exactly the same manner SPIKE-Order and Spike Train Order assign bivariate
order indicators to spikes.
Also these two order indicators, the asymmetric $D_i^{(n,m)}$ and the symmetric
$E_i^{(n,m)}$, which both can take the values $-1$, $0$, or $+1$, are averaged
over spike train pairs and converted into two overall profiles $D(t_k)$ and
$E(t_k)$ which are normalized between $-1$ and $1$.
The SPIKE-Order profile $D(t_k)$ distinguishes leading and following spikes,
whereas the Spike Train Order profile $E(t_k)$ provides information about the
order of spike trains, i.e. it allows to sort spike trains from leaders to
followers. 

First of all, the symmetric coincidence indicator $C^{(n,m)}_i$ of
SPIKE-Synchronization (Eq. \ref{eq:SPIKE-Coincidence}) is replaced by the 
asymmetric SPIKE-Order indicator 
\begin{equation} \label{eq:Bi-SPIKE-Order-1}
	D_i^{(n,m)} = C_i^{(n,m)} \cdot \sign (t_j^{(m)} - t_i^{(n)}),
\end{equation}
where the index $j$ is defined from the minimum
in Eq. \ref{eq:SPIKE-Coincidence} with the threshold-value in Eqs. \ref{eq:A-CoincidenceP}
and \ref{eq:A-CoincidenceF} set to $thr = 0$.

The corresponding value $D_j^{(m,n)}$ is obtained in an asymmetric manner as
\begin{equation} \label{eq:Bi-SPIKE-Order-2}
	D_j^{(m,n)} = C_j^{(m,n)} \cdot \sign (t_i^{(n)} - t_j^{(m)})
		    = - D_i^{(n,m)}.
\end{equation}
Therefore, this indicator assigns to each spike either a $1$ or a $-1$ depending 
on whether the respective spike is leading or following a coincident spike from 
the other spike train.
The value $0$ is obtained for cases in which there is no 
coincident spike in the other spike train ($C^{(n,m)}_i = 0$), but also in cases 
in which the times of the two coincident spikes are absolutely identical  
($t_j^{(m)} = t_i^{(n)}$).

The multivariate profile $D(t_k)$ obtained analogously to Eq. 
\ref{eq:Multi-Profile} is normalized between $1$ and $-1$ and the extreme values 
are obtained if a spike is either leading ($+1$) or following ($-1$) coincident 
spikes in all other spike trains. It can be $0$ either if a spike is not part 
of any coincidences or if it leads exactly as many spikes from other spike 
trains in coincidences as it follows. From the definition in Eqs. 
\ref{eq:Bi-SPIKE-Order-1} and \ref{eq:Bi-SPIKE-Order-2} it follows immediately 
that $C_k$ is an upper bound for the absolute value $|D_k|$.

%
%
\begin{figure}
    \includegraphics[width=85mm]{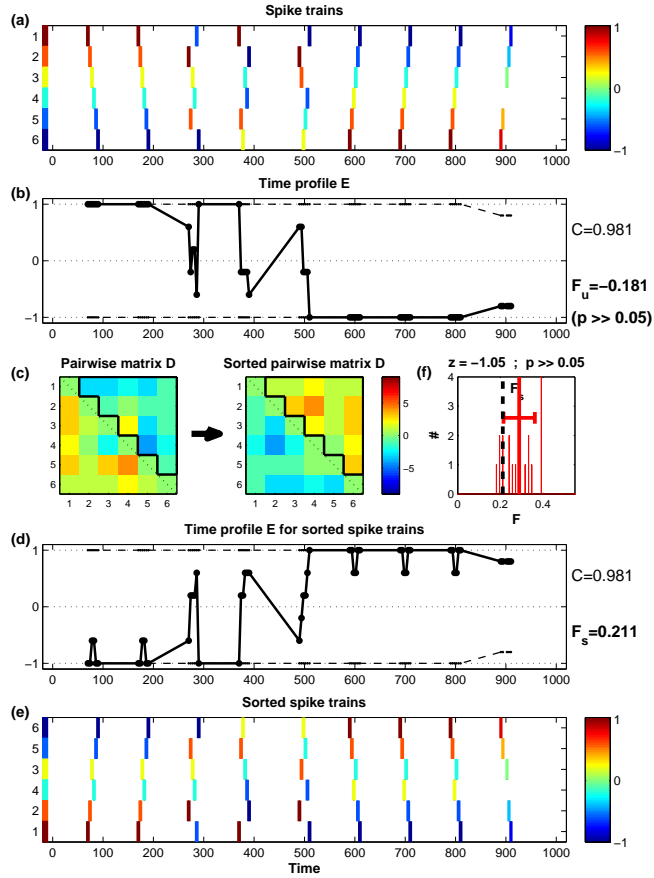}
	\centering
	\caption{\label{fig:SPIKE_Order_Standard_Example}
		Illustration of SPIKE-Order for an artificially created example dataset
		consisting of $6$ spike trains which emit spikes in nine reliable events.
		For the first two events spikes fire in order, for the next three events the
		order is random whereas for the last four events the order is inverted.
		In the last event there is one spike missing.
		(a) Unsorted spike trains with the spikes color-coded according to the value of
		the SPIKE-Order $D(t_k)$.
		(b) Spike Train Order profile $E(t_k)$.
		Events with different firing order can clearly be distinguished.
		The SPIKE-Synchronization profile $C(t_k)$ and its mirror profile (dashed
		black lines) act as envelope.
		The Synfire Indicator $F_u$ for the unsorted spike trains is slightly negative
		reflecting the dominance of the inversely ordered events.
		(c) Pairwise cumulative SPIKE-Order matrix $D$ before (left) and after (right)
		sorting.
		The optimal order maximizes the upper triangular matrix $D_{(n<m)}$, marked in
		black.
		The thick black arrow in between the two matrices indicates the sorting process.
		(d) Spike Train Order profile $E(t_k)$ and its average values, the Synfire
		Indicator $F_s$ for the sorted spike trains.
		(e) Sorted spike trains.
		(f) Statistical significance: Results of the surrogate analysis.
		Thick lines denote mean and standard deviation for $19$ surrogates.
		Since the value for the original dataset (black) is not maximum, the optimally
		sorted spike trains do not exhibit a statistically significant synfire pattern.
}
\end{figure}
%
%
In Fig. \ref{fig:SPIKE_Order_Standard_Example} we show the application of the SPIKE-Order
framework to an example dataset.
While the SPIKE-Order profile can be very useful for color-coding and
visualizing local spike leaders and followers (Fig. 
\ref{fig:SPIKE_Order_Standard_Example}a), it is not useful as an 
overall indicator of Spike Train Order.
The profile is invariant under exchange of spike trains, i.e. it looks the same
for all events no matter what the order of the firing is (in our example only
the last event looks slightly different since one spike is missing).
Moreover, summing over all profile values, which is equivalent to summing over
all coincidences, necessarily leads to an average value of $0$, since for every
leading spike ($+1$) there has to be a following spike ($-1$). 

So in order to quantify any kind of leader-follower information between spike
trains we need a second kind of order indicator.
The Spike Train Order indicator is similar to the SPIKE-Order indicator defined
in Eqs. \ref{eq:Bi-SPIKE-Order-1} and \ref{eq:Bi-SPIKE-Order-2} but with two
important differences.
Both spikes are assigned the same value and this value now depends on the
order of the spike trains:
\begin{equation}  \label{eq:Bi-Spike-train-order-1}
 E_i^{(n,m)} = C_i^{(n,m)} \cdot
 				\begin{cases}
 					\sign (t_j^{(m)} - t_i^{(n)})\quad\text{if}\quad n<m\\
                		\sign (t_i^{(n)} - t_j^{(m)})\quad\text{if}\quad n>m
              	\end{cases}
\end{equation}
and
\begin{equation} \label{eq:Bi-Spik-train-order-2}
	E_j^{(m,n)} = E_i^{(n,m)}.
\end{equation}

This symmetric indicator assigns to both spikes a $+1$ in case the two spikes
are in the correct order, i.e. the spike from the spike train with the lower
spike train index is leading the coincidence, and a $-1$ in the opposite case.
Once more the value $0$ is obtained when there is no coincident spike in the other
spike train or when the two coincident spikes are absolutely identical.

The multivariate profile $E(t_k)$, again obtained similarly to
Eq.~\ref{eq:Multi-Profile}, is also normalized between $1$ and $-1$ and 
the extreme values are obtained for a coincident event covering all spike trains 
with all spikes emitted in the order from first (last) to last (first) spike 
train, respectively (see the first two and the last four events in Fig.
\ref{fig:SPIKE_Order_Standard_Example}).
It can be $0$ either if a spike is not a part of any coincidences or if the
order is such that correctly and incorrectly ordered spike train pairs cancel
each other.
Again, $C_k$ is an upper bound for the absolute value of $E_k$.

\subsection{\label{ss:Synfire-Indicator} Synfire Indicator}

In contrast to the SPIKE-Order profile $D_k$, for the Spike Train Order profile
$E_k$ it does make sense to define an average value, which we term the Synfire
Indicator:
\begin{equation} \label{eq:Synfire Indicator-E}
	F = \frac{1}{M} \sum_{k=1}^M E(t_k).
\end{equation}

The interpretation is very intuitive.
The Synfire Indicator $F$ quantifies to what degree the spike trains in their
current order resemble a perfect synfire pattern.
It is normalized between $1$ and $-1$ and attains the value $1$ ($-1$) if the
spike trains in their current order form a perfect (inverse) synfire pattern,
meaning that all spikes are coincident with spikes in all other spike trains and
that the order from leading (following) to following (leading) spike train
remains consistent for all of these events.
It is $0$ either if the spike trains do not contain any coincidences at all or
if among all spike trains there is a complete symmetry between leading and
following spikes. 

The Spike Train Order profile $E(t_k)$ for our example is shown in Fig. 
\ref{fig:SPIKE_Order_Standard_Example}c.
In this case the order of spikes within an event clearly matters.
The Synfire Indicator $F$ is slightly negative indicating that the current
order of the spike trains is actually closer to an inverse synfire pattern.

Given a set of spike trains we now would like to sort the spike trains from
leader to follower such that the set comes as close as possible to a synfire
pattern.
To do so we have to maximize the overall number of correctly ordered
coincidences and this is equivalent to maximizing the Synfire Indicator $F$.
However, it would be very difficult to achieve this maximization by means of
the multivariate profile $E(t_k)$.
Clearly, it is more efficient to sort the spike trains based on a pairwise
analysis of the spike trains.
The most intuitive way is to use the anti-symmetric cumulative SPIKE-Order
matrix
\begin{equation}
    D^{(n,m)} = \sum_i D_i^{(n,m)}
\end{equation}
which sums up orders of coincidences from the respective pair of spike trains 
only and quantifies how much spike train $n$ is leading spike train $m$ (Fig. 
\ref{fig:SPIKE_Order_Standard_Example}c). 

Hence if $D^{(n,m)}>0$ spike train $n$ is leading $m$, while $D^{(n,m)}<0$ means 
$m$ is leading $n$.
If the current Spike Train Order is consistent with the synfire property, we 
thus expect that $D^{(n,m)} > 0$ for $n<m$ and $D^{(n,m)} < 0$ for $n>m$. 
Therefore, we construct the overall SPIKE-Order as
\begin{equation} \label{eq:synfire_ind_D}
 	D_{(n<m)} = \sum_{n<m} D^{(n,m)},
\end{equation}
i.e.\ the sum over the upper right tridiagonal part of the matrix $D^{(n,m)}$.

After normalizing by the overall number of possible coincidences, we arrive at
a second more practical definition of the Synfire Indicator:
\begin{equation} \label{eq:Synfire Indicator-D}
	F = \frac{2 D_{(n<m)}}{(N-1) M}. 
\end{equation}
The value is identical to the one of Eq. \ref{eq:Synfire Indicator-E}, only
the temporal and the spatial summation of coincidences (i.e., over the profile
and over spike train pairs) are performed in the opposite order.

Having such a quantification depending on the order of spike trains, we can 
introduce a new ordering in terms of the spike train index permutation $\varphi(n)$.
The overall Synfire Indicator for this permutation is then denoted as
$F_\varphi$.
Accordingly, for the initial (\textbf{u}nsorted) order of spike trains 
$\varphi_u$ the Synfire Indicator is denoted as $F_u = F_{\varphi_u}$.

The aim of the analysis is now to find the optimal (\textbf{s}orted) order
$\varphi_s$ as the one resulting in the maximal overall Synfire Indicator
$F_s = F_{\varphi_s}$:
\begin{equation}
	\varphi_s: F_{\varphi_s} = \max_\varphi \{F_\varphi\} = F_s.
\end{equation}
This Synfire Indicator for the sorted spike trains quantifies how close spike
trains can be sorted to resemble a synfire pattern, i.e., to what extent
coinciding spike pairs with correct order prevail over coinciding spike pairs
with incorrect order.
Unlike the Synfire Indicator for the unsorted spike trains $F_u$, the optimized
Synfire Indicator $F_s$ can only attain values between $0$ and $1$ (any order that
yields a negative result could simply be reversed in order to obtain the same
positive value).
For a perfect synfire pattern we obtain $F_s=1$, while sufficiently long
Poisson spike trains without any synfire structure yield $F_s\approx 0$.

The complexity of the problem to find the optimal Spike Train Order is 
similar to the well-known travelling salesman problem \cite{Applegate11}. 
For $N$ spike trains there are $N!$ permutations~$\varphi$, so for large numbers
of spike trains finding the optimal Spike Train Order $\varphi_s$ is a non-trivial
problem and brute-force methods such as calculating the $F_\varphi$-value for 
all possible permutations are not feasible. 
Instead, we search for the optimal order using simulated annealing \cite{Dowsland12},
a probabilistic technique which approximates the global optimum of a given function
in a large search space.
In our case this function is the Synfire Indicator $F_\varphi$ (which we would
like to maximize) and the search space is the permutation space of all spike
trains.
We start with the $F_u$-value from the unsorted permutation and then visit
nearby permutations using the fundamental move of exchanging two neighboring
spike trains within the current permutation. 
All moves with positive $\Delta F$ are accepted while the likelihood of 
accepting moves with negative $\Delta F$ is decreased along the way according to 
a standard slow cooling scheme.
The procedure is repeated iteratively until the order of the spike trains no
longer changes or until a predefined end temperature is reached.

The sorting of the spike trains maximizes the Synfire Indicator as reflected by
both the normalized sum of the upper right half of the pairwise cumulative
SPIKE-Order matrix (Eq.\ref{eq:Synfire Indicator-D}, Fig.
\ref{fig:SPIKE_Order_Standard_Example}c) and the average value of the
Spike Train Order profile $E (t_k)$ (Eq.\ref{eq:Synfire Indicator-E},
Fig. \ref{fig:SPIKE_Order_Standard_Example}d).
Finally, the sorted spike trains in Fig. \ref{fig:SPIKE_Order_Standard_Example}e
are now ordered such that the first spike trains have predominantly high values
(red) and the last spike trains predominantly low values (blue) of $D (t_k)$.

The complete analysis returns results consisting of several levels of
information.
Time-resolved (local) information is represented in the spike-coloring and in
the profiles $D$ and $E$.
The pairwise information in the SPIKE-order matrix reflects the leader-follower
relationship between two spike trains at a time.
The Synfire Indicator $F$ characterizes the closeness of the dataset as a whole
to a synfire pattern, both for the unsorted ($F_u$) and for the sorted ($F_s$)
spike trains.
Finally, the sorted order of the spike trains is a very important result in
itself since it identifies the leading and the following spike trains.

\subsection{\label{ss:Statistical-Significance} Statistical significance}

As a last step in the analysis we evaluate the statistical significance of the 
optimized Synfire Indicator $F_s$ using a set of carefully constructed spike 
train surrogates. The idea behind the surrogate test is to estimate the 
likelihood that the consistent SPIKE-Order pattern yielding a certain Synfire 
Indicator could have been obtained by chance. To this aim, for each surrogate we 
maintain the coincidence structure of the spike trains by keeping the 
SPIKE-Synchronization values of each individual spike constant but randomly swap 
the spike order in a sufficient number of coincidences. We set the number of 
swaps equal to the number of coincident spikes in the dataset since this way all 
possible spike order patterns can be reached. Only for the first surrogate we 
swap twice as many coincidences in order to account for transients. After each 
swap we take extra care that all other spike orders that are affected by the 
swap are updated as well. For example, if a swap changes the order between the 
first and the third spike in an ordered sequence of three spikes, we also swap 
both the order between the first and the second and the order between the second 
and the third spike.

For each spike train surrogate we repeat exactly the same optimization procedure 
in the spike train permutation space that is done for the original dataset.
The original Synfire Indicator is deemed significant if it is higher than the 
Synfire Indicator obtained for all of the surrogate datasets (this case will be 
marked by two asterisks).
Here we use $s = 19$ surrogates for a significance level of
$p^* = 1/(s+1) = 0.05$.
Note that in order to achieve a better sampling of the underlying null distribution a larger number of surrogates would be preferable but the chosen value of $s$ is a compromise that takes into account the computational cost.
As a second indicator we state the z-score, e.g., the deviation of the original
value $x$ from the mean $\mu$ of the surrogates in units of their standard
deviation $\sigma$:
\begin{equation}
    z = \frac{x-\mu}{\sigma}.
\end{equation}

Results of the significance analysis for our standard example are shown in the 
histogram in Fig. \ref{fig:SPIKE_Order_Standard_Example}f.
In this case the absolute value of the z-score is smaller than one and the
p-value is larger than $p^*$ and the result is thus judged as statistically
non-significant. 

\section{Outlook}
\label{sec:4}

In the first part of this chapter we describe three parameter-free and time resolved measures of spike train synchrony in their recently developed
adaptive extensions, A-ISI-distance, A-SPIKE-distance and A-SPIKE-synchronization
\cite{Satuvuori17}.
All of these measures are symmetric and so their multivariate versions are invariant
to changes in the order of spike trains.
Since information about directionality is very relevant, in the second part of this chapter
we show an algorithm which allows to sort multiple spike trains from leader to follower.
This algorithm is built on two indicators, SPIKE-Order and Spike Train Order, that define
the Synfire Indicator value, which quantifies the consistency of the temporal leader-follower
relationships for both the original and the optimized sorting.

Symmetric measures of spike train distances (Section \ref{sec:2}) have been applied in many
different contexts, not only in the field of neuroscience \cite{Jolivet08, Mainen95, Victor05}.
For example, they have been used in robotics \cite{Espinal16} and prosthesis control
\cite{Dura-Bernal16}.
The ISI-distance has been applied in a method for the detection of directionl coupling
between point processes and point processes and flows \cite{Andrzejak11}, as an
adaptation of the nonlinear technique for directional coupling detection of continuous
signals \cite{Chicharro09b}.  

Questions about leader-follower dynamics (Section \ref{sec:3}) have been specifically investigated
in neuroscience \cite{Pereda05}, but also in fields as wide-ranging as, e.g.,
climatology \cite{Boers14}, social communication \cite{Varni10}, and human-robot
interaction \cite{Rahbar15}.
SPIKE-Order has already been  applied to analyze the consistency of propagation
patterns in two real datasets from neuroscience (Giant Depolarized Potentials in
mice slices) and climatology (El Ni\~no sea surface temperature
recordings) \cite{Kreuz2016}. 

Finally, we would like to mention that the similarity measures A-ISI-distance,
A-SPIKE-distance and A-SPIKE-synchronization, as well as SPIKE-Order, are implemented
in three publicly available software packages, the Matlab-based graphical user interface
SPIKY\footnote{http://www.fi.isc.cnr.it/users/thomas.kreuz/Source-Code/SPIKY.html} \cite{Kreuz15},
cSPIKE\footnote{http://www.fi.isc.cnr.it/users/thomas.kreuz/Source-Code/cSPIKE.html}
(Matlab command line with MEX-files), and the open-source Python library
PySpike\footnote{http://mariomulansky.github.io/PySpike} \cite{Mulansky16}.

\begin{acknowledgement}
We acknowledge funding from the European Union's Horizon 2020
research and innovation program under the Marie Skłodowska-Curie Grant Agreement
No. \#642563 'Complex Oscillatory Systems: Modeling and Analysis' (COSMOS).
T.K. also acknowledges support from the European Commission through Marie Curie Initial
Training Network 'Neural Engineering Transformative Technologies' (NETT), project
289146.
We thank Ralph G. Andrzejak, Nebojsa Bozanic, Kerstin Lenk, Mario Mulansky, and Martin Pofahl for
useful discussions.
\end{acknowledgement}

\bibliographystyle{plain}
\bibliography{Satuvuori_Malvestio_Kreuz_2017}

\end{document}